\documentclass[12pt]{iopart}
\pdfoutput=1
\usepackage{bm}
\usepackage{amssymb}
\usepackage{pifont}
\usepackage{slashed} 
\usepackage{slantsc} 
\usepackage[usenames,dvipsnames,svgnames,table]{xcolor}

\begin{document}
\title{%
Cartesian Kerr--Schild variation on the Newman--Janis trick
}
\author{Del Rajan \hbox{{\sf and}} Matt Visser\,}
\address{School of Mathematics and Statistics,
Victoria University of Wellington; \\
PO Box 600, Wellington 6140, New Zealand.}
\begin{abstract}
The Newman--Janis trick is a procedure, (not even really an ansatz), for obtaining the Kerr spacetime from the Schwarzschild spacetime. This 50 year old trick continues to generate heated discussion and debate even to this day. Most of the debate focusses on whether the Newman--Janis procedure can be upgraded to the status of an algorithm, or even an inspired ansatz, or is it just a  random trick of no deep physical significance. (That the Newman--Janis procedure very quickly led to the discovery of the Kerr--Newman spacetime is a point very much in its favour.) In the current article we will not answer these deeper questions, we shall instead present a much simpler alternative variation on the theme of the Newman--Janis trick that might be easier to work with. We shall  present a 2-step version of the Newman--Janis trick that works directly with the Kerr--Schild ``Cartesian'' metric presentation of the Kerr spacetime.  
That is, we show how the original 4-step Newman--Janis procedure can, (using the interplay between oblate spheroidal and Cartesian coordinates), be reduced to a considerably cleaner 2-step process. 

\medskip
\noindent{\it Keywords\/}:
Newman--Janis trick; ansatz; algorithm; Kerr spacetime;  Kerr--Newman spacetime;
Schwarzschild spacetime. arXiv: 1601.03532 [gr-qc]

\medskip
\noindent
D{\sc{ate}}: 15 February 2017; \LaTeX-ed \today.

\end{abstract}

\pacs{04.20.-q; 04.20.Cv; 04.62.+v; 04.70.-s}

\maketitle
\section{Introduction}

The Newman--Janis trick was formulated some 50 years ago~\cite{Newman:1965a}, immediately leading to the discovery of the Kerr--Newman spacetime~\cite{Newman:1965b} describing electrically charged rotating black holes. Over the last 50 years there has been continual and sometimes heated debate and discussion as to the status of this trick --- can the trick be upgraded to an algorithm, or at least an ansatz, or is it essentially just a lucky guess. Part of the reason for the 50 years of confusion is that there is some considerable ambiguity as to what exactly the Newman--Janis procedure actually is; as several steps in the Newman--Janis procedure involve some seemingly arbitrary choices whose physical justification seems somewhat less than pellucid~\cite{MSc}. 
For general background discussion see for instance~\cite{Newman:1973, Drake:1998, Kerr:2007, Whisker:2008, Adamo:2009, Adamo:2014, Ferraro:2013, Keane:2014, Harte:2014, Erbin:2014a, Erbin:2014b, Erbin:2015, Harte:2016, Erbin:2017}.

Now the original Newman--Janis trick is a 4-step procedure that works with complex orthonormal null tetrads. While there are certainly good physics reasons to believe that all physically interesting Lorentzian spacetimes always admit globally defined orthonormal null tetrads~\cite{global},  both the Schwarzschild and Kerr spacetimes can be described directly in terms of the metric, without introducing null tetrads. This hints that it might be possible to find a variant of the Newman--Janis trick that avoids null tetrads altogether. One such variant is Giampieri's version of the trick~\cite{Giampieri}. Another variant, described below, is a much simpler 2-step procedure that works by using the interplay between oblate spheroidal and Cartesian coordinates to motivate the Kerr--Schild ``Cartesian'' form of the Kerr metric. Both of these metric-based  variants of the Newman--Janis trick still require one to make some seemingly arbitrary choices, but we would argue that they are somewhat better motivated than the original Newman--Janis trick.

\section{Kerr--Schild Cartesian form of the Newman--Janis trick}

Consider ordinary spherical polar coordinates $(r_0,\theta_0,\phi_0)$ on flat 3-space
\begin{equation}
x+iy =   r_0\;\sin\theta_0 \; e^{i\phi_0}, 
\qquad
z = r_0 \;\cos\theta_0,
\end{equation}
and compare them with oblate spheroidal coordinates $(r,\theta,\phi)$ on the same flat 3-space:
\begin{equation}
x+iy =    (r+ia)\;\sin\theta \; e^{i\phi}, 
\qquad
z = r \;\cos\theta.
\end{equation}
In Cartesian coordinates, the Schwarzschild geometry in Kerr--Schild form is
\begin{equation}
g_{ab} = \eta_{ab} + {2m\over r_0} \; (\ell_0)_a \; (\ell_0)_b; 
\end{equation}
where $\eta_{ab} = \mathrm{diag}\{-1,+1,+1,+1\}$ and 
\begin{equation}
(\ell_0)_a = (1; \hat r) =  \left( 1; {x\over r_0}, {y\over r_0}, {z\over r_0}\right) = 
 \left( 1; \sin\theta_0\cos\phi_0, \sin\theta_0\sin\phi_0, \cos\theta_0  \right).
\end{equation} 

\clearpage
\noindent
To go from Schwarzschild to Kerr we now simply do two (quite independent) things:
\begin{enumerate}
\item 
Replace
\begin{eqnarray}
(\ell_0)_a &=&  \left( 1; \sin\theta_0\cos\phi_0, \sin\theta_0\sin\phi_0, \cos\theta_0  \right) 
\nonumber\\
&\longrightarrow&
\left( 1; \sin\theta\cos\phi, \sin\theta\sin\phi, \cos\theta  \right) = \ell_a.
\label{E:pedestrian}
\end{eqnarray}
\item 
Replace
\begin{equation}
{2m\over r_0} \longrightarrow  m\left({1\over r+ia\cos\theta} + {1\over r-ia\cos\theta} \right) = {2mr\over r^2+a^2\cos^2\theta}.
\end{equation}
\end{enumerate}
 That's all; no other changes are required. This is the fastest and most direct way we know of for going from Schwarzschild to Kerr. 
(In particular we do not need to invert the metric, nor do we need to work with a complex null tetrad, nor are any coordinate changes [whether real or complex] involved.) The original 4-step Newman--Janis procedure~\cite{Newman:1965a} has now been reduced to a 2-step process.

\section{Verifying correctness of this version of the  trick}

To check that this procedure actually does generate the Kerr spacetime we proceed as follows.
\begin{itemize}
\item 
First, we note:
\begin{eqnarray}
\ell_a &=&  \left( 1; \sin\theta\cos\phi, \sin\theta\sin\phi, \cos\theta  \right)
\nonumber\\
&=& \left( 1; \Re\left[{x+iy\over r+ia}\right], \Im\left[{x+iy\over r+ia}\right], {z\over r}  \right)
\nonumber\\
&=& \left( 1; \Re\left[(x+iy)(r-ia)\over r^2+a^2\right], \Im\left[(x+iy)(r-ia)\over r^2+a^2\right], {z\over r}  \right)
\nonumber\\
&=& \left( 1; {xr+ay\over r^2+a^2}, {yr-ax\over r^2+a^2}, {z\over r}  \right).
\end{eqnarray}
\item
Second, we observe:
\begin{eqnarray}
x^2+y^2+z^2 &=& (r^2+a^2)\sin^2\theta + r^2 \cos^2\theta 
\nonumber\\
&=&
 r^2 + a^2\sin^2\theta 
 \nonumber\\
&=& r^2+a^2 \left[1-{z^2\over r^2}\right].
\end{eqnarray}
Explicitly, using $r_0^2 = x^2+y^2+z^2$, we have
\begin{equation}
r^2 = {1\over2} \left( r_0^2 - a^2 + \sqrt{(r_0^2 - a^2)^2 +4z^2 a^2 } \right).
\end{equation}
\item
Third, we ascern:
\begin{equation}
{2mr\over r^2+a^2\cos^2\theta} =  {2mr^3\over r^4 +a^2 z^2}.
\end{equation}
\end{itemize}
These three equations are exactly the defining equations for the Kerr space-time in Cartesian Kerr--Schild form. 
See any standard textbook or review, for instance~\cite{Hawking-Ellis, Sachs-Wu, toolkit, Visser:2007, Johannsen:2015}.

\clearpage
\noindent
That is:
\begin{equation}
g_{ab} = \eta_{ab} + {2mr\over r^2+a^2\cos^2\theta}  \; \ell_a \ell_b
\end{equation}
is Ricci flat. Why does this work? As always with the Newman--Janis procedure and its variants, this ``why'' is the central question.

\section{Why does the Newman--Janis trick work?}

There are two separate mappings involved in this procedure, the transformation of the null covector $(\ell_0)_a\longrightarrow \ell_a$, and the transformation $f_0(x) \to f(x)$ of the scalar prefactor  in the Kerr--Schild metric 
$(g_0)_{ab} = \eta_{ab} + f_0(x)  \; (\ell_0)_a (\ell_0)_b \;\;\longrightarrow\;\; g_{ab} = \eta_{ab} + f(x)  \; \ell_a \ell_b$. 
We \emph{motivate} these transformations, (we do not \emph{derive} these transformations), below.

\paragraph{Regarding the null covector:} With hindsight, part of the  motivation for the trick is obvious --- there has been over 200 years of work on rotating fluid bodies in Newtonian gravity, and for close to 200 years it has been clear that (flat space) oblate spheroidal coordinates have been useful in this regard --- so it should not really be a surprise that oblate spheroidal coordinates also show up in rotating black holes. 

What is, however, surprising is the extreme and quite unexpected simplicity of the mapping $(\ell_0)_a\to \ell_a$ taking the Schwarzschild radial null geodesic congruence $(\ell_0)_a$, to Kerr's null geodesic congruence $\ell_a$:
\begin{equation}
\fl\qquad\qquad
\left( 1; \sin\theta_0\cos\phi_0, \sin\theta_0\sin\phi_0, \cos\theta_0  \right) 
\longrightarrow
 \left( 1; \sin\theta\cos\phi, \sin\theta\sin\phi, \cos\theta  \right).
\end{equation}
This amounts to just dropping the subscript zero and so ``reinterpreting'' spherical polar coordinates in terms of oblate spheroidal coordinates. 
In terms of the underlying Cartesian coordinates this corresponds to
\begin{equation}
\left( 1; {x\over r_0}, {y\over r_0}, {z\over r_0}\right)
\longrightarrow
\left( 1; {xr+ay\over r^2+a^2}, {yr-ax\over r^2+a^2}, {z\over r}  \right).
\end{equation}
This is not merely a coordinate transformation, it is a genuine change in the null vector. While it is clear that this procedure automatically yields a null congruence, it is not \emph{a~priori} clear that this congruence will be geodesic, let alone shear-free.  That can only be checked \emph{a posteriori}.  (See for instance reference~\cite{Ayon-Beato:2015}.)

If, instead of the covariant vector $\ell_a$, we choose to work with the contravariant null vector $\ell^a = \eta^{ab} \,\ell_b$, we obtain closely related results. Starting from
\begin{equation}
\fl
\partial_r 
= (\partial_r x) \partial_x + (\partial_r y) \partial_y+(\partial_r z) \partial_z 
=(\sin\theta\cos\phi) \partial_x + (\sin\theta\sin\phi) \partial_y+(\cos\theta) \partial_z,
\end{equation}
we see that
\begin{equation}
\fl
(\ell_0)^\sharp = (\ell_0)^a \partial_a = -\partial_t + (\sin\theta_0\cos\phi_0) \partial_x + (\sin\theta_0\sin\phi_0) \partial_y+(\cos\theta_0) \partial_z 
= -\partial_t + \partial_{r_0},
\end{equation}
whereas
\begin{equation}
\fl
(\ell)^\sharp = \ell^a \partial_a = -\partial_t + (\sin\theta\cos\phi) \partial_x + (\sin\theta\sin\phi) \partial_y+(\cos\theta) \partial_z 
= -\partial_t + \partial_{r}.
\end{equation}
So for the covariant vector transformation $(\ell_0)^\sharp \to \ell^\sharp$ we simply need to consider
\begin{equation}
(\ell_0)^\sharp  = -\partial_t + \partial_{r_0} \quad\longrightarrow\quad (\ell)^\sharp = - \partial_t + \partial_{r}.
\end{equation}
This amounts to just dropping the subscript zero and ``reinterpreting'' spherical polar coordinate $r_0$ in terms of the oblate spheroidal coordinate $r$. 
If we furthermore introduce retarded times $u_0 = t-r_0$ and $u=t-r$, while keeping other coordinates fixed, then this becomes even simpler:
\begin{equation}
(\ell_0)^\sharp  = -\partial_{u_0} \quad\longrightarrow\quad (\ell)^\sharp = - \partial_u.
\end{equation}

\paragraph{Regarding the Kerr--Schild scalar potential prefactor $f(x)$:}
For this half of the Newman--Janis trick a clue comes from the well-known behaviour of the Schwarzschild potential term $2m/r$ under complex displacement of the source. (See for instance references~\cite{Schiffer:1973, Santos}.) Formally move the source in the Schwarzschild solution from the origin to $i a \hat z$, then in spheroidal coordinates we have
\begin{equation}
\fl\qquad\quad
{1\over r_0} \longrightarrow {1\over r_c} = {1\over\sqrt{x^2+y^2+(z-ia)^2}} = {1\over\sqrt{(r^2+a^2)\sin^2\theta +(r \cos\theta -ia)^2}}.
\end{equation}
But this simplifies to
\begin{equation}
{1\over r_c} = {1\over\sqrt{r^2- a^2\cos^2\theta - 2ia r \cos\theta}} = {1\over\sqrt{(r- ia \cos\theta)^2}} = {1\over r -ia\cos\theta} .
\end{equation}
That is, under translation of the source along the imaginary $\hat z$ axis we have 
\begin{equation}
{1\over r_0} \to {1\over r_c} = {1\over r -ia\cos\theta}.
\end{equation}
Complex translations along these (and similar)  lines are also useful in many other situations~\cite{Visser:2003}. 
Another clue comes from the fact that for metrics that can be written in the Kerr--Schild form, $g_{ab}=\eta_{ab}+f(x)\,\ell_a\ell_b$, with $\ell_a$ a shear-free null geodesic congruence, the Ricci tensor is in fact linear in $f(x)$~\cite{Schiffer:1973, Gergely:2002}. This motivates superimposing two half-strength sources at $\pm i a \hat z$, and setting
\begin{equation}
{1\over r_0} \to {1\over 2 r_c} +  {1\over 2 \bar r_c} ={1\over 2(r +ia\cos\theta)}+ {1\over 2(r -ia\cos\theta)} =
{r\over r^2+a^2\cos^2\theta}.
\end{equation}

\paragraph{Cautionary comment:}
 Individually all of these observations seem (at least \emph{a posteriori}) tolerably well motivated, but there seems to be two quite separate things going on: The null congruence mapping  $(\ell_0)_a\to \ell_a$ does not seem to have any nice simple interpretation in terms of translations along the imaginary $\hat z$ axis. A naive attempt along these lines might be to consider
 \begin{eqnarray}
(1; \nabla r_0) = \left( 1; {x\over r_0}, {y\over r_0}, {z\over r_0}\right)
&\quad\longrightarrow\quad& (1;\nabla r_c) = 
\left( 1; {x\over r_c}, {y\over r_c}, {z-ia\over r_c}\right).
\end{eqnarray}
While this is certainly a null vector (with complex components) it is unfortunately not obviously related to Kerr's shear-free null geodesic congruence
\begin{equation}
\ell_a = \left( 1; {xr+ay\over r^2+a^2}, {yr-ax\over r^2+a^2}, {z\over r}  \right).
\end{equation}
With considerable hindsight, the Schiffer--Adler--Mark--Sheffield article~\cite{Schiffer:1973} gives some useful information. 
Their arguments (after a little work) imply, see their equation (4.30), that the Kerr null congruence can be written as:
\begin{equation}
\ell_a = \left(1; {\nabla r_c + \nabla\bar r_c - i \; \nabla r_c \times \nabla\bar r_c\over 
1 +\nabla r_c \cdot\nabla\bar r_c}
\right).
\label{E:complex}
\end{equation}
This real null vector depends \emph{non-homomorphically} and \emph{non-linearly} on the complex distance $r_c$, and its gradient $\nabla r_c$, and has no really straightforward interpretation  in terms of any analytic continuation of the original Schwarzschild null congruence. It is the lack of uniformity in the treatment of the radial coordinate, (sometimes $r_0\to r_c$, sometimes $r_0\to \bar r_c$, sometimes worse as in the cross product term), that is disturbing. 
(The way that Schiffer--Adler--Mark--Sheffield establish the equivalent of this result is quite indirect, by first partially solving the vacuum Einstein equations to obtain a nontrivial nonlinear relationship between the Kerr null congruence $\ell_a$ and the potential  $2m/r_c$.)

\paragraph{Extension \#1:} From the above discussion it is clear that
\begin{equation}
g_{ab} = \eta_{ab} + {2m\over r \pm ia \cos\theta}  \; \ell_a \ell_b
\end{equation}
is also Ricci flat. Note this is a complex-valued metric tensor over a real manifold; this metric is not itself of direct physical interest, but metrics of this type typically do occur at intermediate stages in many computations based on the Newman--Janis algorithm~\cite{MSc}.

\paragraph{Extension \#2:}
What happens if we allow mass to become complex? Then
\begin{equation}
\fl
{2m\over r_0} \to {m\over r_c} + {\bar m \over \bar r_c} = 
{m\over r+ia\cos\theta} + {\bar m\over r-ia\cos\theta} 
= {2\Re(m)r+2\Im(m)a\cos\theta \over r^2+a^2\cos^2\theta}.
\end{equation}
Then automatically
\begin{equation}
g_{ab} = \eta_{ab} + {2\Re(m)r+2\Im(m)a\cos\theta \over r^2+a^2\cos^2\theta}  \; \ell_a \ell_b
\end{equation}
is also Ricci flat. Note this is now a real-valued metric tensor over a real manifold. But the parameter  $\Im(m)$ is now proportional to the NUT charge, and so destroys asymptotic flatness~\cite{Kerr:2007, Erbin:2014b, Erbin:2015, DNJ:1996}.  For this reason it is typically set to zero early in the calculation. 

\paragraph{Extension \#3:}
{The Schwarzschild to Reissner--Nordstr\"om transition corresponds to}
\begin{equation}
{2m\over r_0 } \quad\longrightarrow\quad {2m\over r_0 }  + {q^2\over r_0^2}.
\end{equation}
But the Reissner--Nordstr\"om to Kerr--Newman transition corresponds to
\begin{equation}
{2m\over r_0 }  + {q^2\over r_0^2} \quad\longrightarrow\quad {m\over r_c} + {m\over \bar r_c} + {q^2\over r_c \bar r_c}. 
\end{equation}
It is the lack of uniformity in the treatment of the radial coordinate, (sometimes $r_0\to r_c$, sometimes $r_0\to \bar r_c$), that is again disturbing. 

\paragraph{Extension \#4:}
A reasonably general ansatz is to set
\begin{equation}
g_{ab} = \eta_{ab} + f(r_c, \bar r_c)\; \ell_a \ell_b,
\end{equation}
where $f(r_c, \bar r_c)$ is some real function of its two complex conjugate arguments. This ansatz is sufficent to capture most of the interesting physics.

\section{Discussion}

So what have we actually achieved in this article? We have developed a very simple and direct version of the Newman--Janis trick, one that works directly on the (covariant) metric.  (So we can avoid working with null tetrads, metric inversion, and even avoid complex coordinate transformations.) In doing so we have reduced the usual four-step process to a much more direct two-step process. We do not claim to have upgraded the Newman--Janis procedure to an algorithm, the procedure still requires some motivated guesswork, so it is still at best a trick. 
It is the compact nature of the current two-step version of the Newman--Janis trick that is the primary issue of scientific interest in this article.

Even with this simplified two-step process, there are still two radically different things going on. The complexification of the ``potential'' prefactor is straightforward, and in Newtonian gravity has a history going back some 130 years~\cite{Appell}. 
In contrast the transition from the Schwarzschild null geodesic congruence to the Kerr null geodesic congruence is much trickier. One route, via the deformation of spherical polar to oblate spheroidal coordinates is technically elementary, even pedestrian, but somewhat mysterious.  See equation (\ref{E:pedestrian}). Another route, via the complex distance function $r_c$,  is perhaps more elegant, but if anything even more mysterious. See equation (\ref{E:complex}). Overall, on balance  we still feel that the Newman--Janis procedure is a trick rather than an algorithm.

\section*{Acknowledgments}

This research was supported by the Marsden Fund, through a grant administered by the Royal Society of New Zealand. 

\section*{References}


\begin{thebibliography}{69}

\bibitem{Newman:1965a}
  E.~T.~Newman and A.~I.~Janis,\\
  ``Note on the Kerr spinning particle metric'',\\
  J.\ Math.\ Phys.\  {\bf 6} (1965) 915.\\
  doi:10.1063/1.1704350

\bibitem{Newman:1965b}
  E.~T.~Newman, R.~Couch, K.~Chinnapared, A.~Exton, A.~Prakash and R.~Torrence,\\
  ``Metric of a rotating, charged mass'',\\
  J.\ Math.\ Phys.\  {\bf 6} (1965) 918.\\
  doi:10.1063/1.1704351
  

 \bibitem{MSc}
 Del Rajan, \\
 ``Complex spacetimes and the Newman--Janis trick'',\\
 MSc Thesis, Victoria University of Wellington, 2015. 
 

 \bibitem{Newman:1973}
  ET Newman,\\
  ``Complex coordinate transformations and the Schwarzschild--Kerr metrics'',\\
  Journal of Mathematical Physics {\bf 14(6)} (1973) 774--776.
  
  \bibitem{Drake:1998}
  S.~P.~Drake and P.~Szekeres,\\
  ``Uniqueness of the Newman--Janis algorithm in generating the Kerr--Newman metric'',\\
  Gen.\ Rel.\ Grav.\  {\bf 32} (2000) 445\\
  doi:10.1023/A:1001920232180\\{}
  [gr-qc/9807001].


\bibitem{Kerr:2007}
  R.~P.~Kerr,\\
  ``Discovering the Kerr and Kerr--Schild metrics'',\\
  arXiv:0706.1109 [gr-qc].\\
  Published in \emph{The Kerr spacetime: Rotating black holes in general relativity},\\
  edited by David Wiltshire, Matt Visser, and Susan Scott.\\
  Cambridge University Press, Cambridge, 2009.
  
    \bibitem{Whisker:2008}
  R.~Whisker,\\
  ``Braneworld Black Holes'',\\
  PhD Thesis, Durham,\\
  arXiv:0810.1534 [gr-qc].

  
  \bibitem{Adamo:2009}
  T.~M.~Adamo, C.~N.~Kozameh and E.~T.~Newman,\\
  ``Null geodesic congruences, asymptotically flat space-times and their physical interpretation'',\\
  Living Rev.\ Rel.\  {\bf 12} (2009) 6\\{}
   [Living Rev.\ Rel.\  {\bf 15} (2012) 1]\\{}
  [arXiv:0906.2155 [gr-qc]].
  
  \bibitem{Adamo:2014}
  T.~Adamo and E.~T.~Newman,\\
  ``The Kerr--Newman metric: A review'',\\
  Scholarpedia {\bf 9} (2014) 31791\\
  doi:10.4249/scholarpedia.31791\\{}
  [arXiv:1410.6626 [gr-qc]].
  
  \bibitem{Ferraro:2013}
  R.~Ferraro,\\
  ``Untangling the Newman--Janis algorithm'',\\
  Gen.\ Rel.\ Grav.\  {\bf 46} (2014) 1705\\
  doi:10.1007/s10714-014-1705-3\\{}
  [arXiv:1311.3946 [gr-qc]].
  
  \bibitem{Keane:2014}
  A.~J.~Keane,\\
  ``An extension of the Newman--Janis algorithm'',\\
  Class.\ Quant.\ Grav.\  {\bf 31} (2014) 155003\\
  doi:10.1088/0264-9381/31/15/155003\\{}
  [arXiv:1407.4478 [gr-qc]].
  
\bibitem{Harte:2014}
  A.~I.~Harte,\\
  ``Taming the nonlinearity of the Einstein equation'',\\
  Phys.\ Rev.\ Lett.\  {\bf 113} (2014) no.26,  261103\\
  doi:10.1103/PhysRevLett.113.261103\\{}
  [arXiv:1409.4674 [gr-qc]].

  
  \bibitem{Erbin:2014a}
  H.~Erbin,\\
   ``Deciphering and generalizing Demia\'nski--Janis--Newman algorithm'',\\
  Gen.\ Rel.\ Grav.\  {\bf 48} (2016) 56\\
  doi:10.1007/s10714-016-2054-1\\{}
  [arXiv:1411.2909 [gr-qc]].

  
  \bibitem{Erbin:2014b}
  H.~Erbin and L.~Heurtier,\\
  ``Five-dimensional Janis--Newman algorithm'',\\
  Class.\ Quant.\ Grav.\  {\bf 32} (2015) 16,  165004\\
  doi:10.1088/0264-9381/32/16/165004\\{}
  [arXiv:1411.2030 [gr-qc]].
  
  \bibitem{Erbin:2015}
  H.~Erbin and L.~Heurtier,\\
  ``Supergravity, complex parameters and the Janis--Newman algorithm'',\\
  Class.\ Quant.\ Grav.\  {\bf 32} (2015) 165005\\
  doi:10.1088/0264-9381/32/16/165005\\{}
  [arXiv:1501.02188 [hep-th]].
  
   \bibitem{Harte:2016}
  A.~I.~Harte and J.~Vines,\\
  ``Generating exact solutions to Einstein's equation using linearized approximations'',\\
  Phys.\ Rev.\ D {\bf 94} (2016) no.8,  084009\\
  doi:10.1103/PhysRevD.94.084009\\{}
  [arXiv:1608.04359 [gr-qc]].

  \bibitem{Erbin:2017}
  H.~Erbin,\\
  ``Janis--Newman algorithm: Generating rotating and NUT charged black holes'',\\
  arXiv:1701.00037 [gr-qc].
  
  
 
 \bibitem{global}
 Del Rajan and Matt Visser, \\
 ``Global properties of physically interesting Lorentzian spacetimes'',\\
 International Journal of Modern Physics {\bf D25} (2016) 1650106;\\
doi: 10.1142/S0218271816501066\\{}
  [arXiv:1601.03355 [gr-qc]].


\bibitem{Giampieri}
Giacomo Giampieri,\\
``Introducing angular momentum into a black hole using complex variables'',\\
Gravity Research Foundation essay contest, 1990. \\
{\sf gravityresearchfoundation.org/pdf/awarded/1990/giampieri.pdf}



\bibitem{Hawking-Ellis}
 SW Hawking and GFR Ellis,\\
 \emph{The large scale structure of spacetime},\\
 Cambridge University Press, Cambridge, 1973.


\bibitem{Sachs-Wu}
RK Sachs and H Wu,\\
\emph{General relativity for mathematicians},\\
Springer--Verlag, New York, 1977.

\bibitem{toolkit}
Eric Poisson,\\
\emph{A relativist's toolkit},\\
Cambridge University Press, Cambridge, 2004.\\
ISBN-13: 978-0521537803\;\;
ISBN-10: 0521537800

\bibitem{Visser:2007}
  M.~Visser,\\
  ``The Kerr spacetime: A brief introduction'',\\
  arXiv:0706.0622 [gr-qc].\\
  Published in \emph{The Kerr spacetime: Rotating black holes in general relativity},\\
  edited by David Wiltshire, Matt Visser, and Susan Scott.\\
  Cambridge University Press, Cambridge, 2009.
  
  \bibitem{Johannsen:2015}
  T.~Johannsen,\
  ``Sgr A* and General Relativity'',\\
  Class.\ Quant.\ Grav.\  {\bf 33} (2016) 113001\\
  doi:10.1088/0264-9381/33/11/113001\\{}
  [arXiv:1512.03818 [astro-ph.GA]].


\clearpage

\bibitem{Ayon-Beato:2015}
  E.~Ay\'on-Beato, M.~Hassa\"ine and D.~Higuita-Borja,\\
  ``The role of symmetries in the Kerr--Schild derivation of the Kerr black hole'',\\
  arXiv:1512.06870 [hep-th].
  




\bibitem{Schiffer:1973}
Menahem M. Schiffer, Ronald J. Adler, James Mark, and Charles Sheffield,\\
``Kerr geometry as complexified Schwarzschild geometry'',\\
Journal of Mathematical Physics {\bf 14}  (1973) 52--56\\
http://dx.doi.org/10.1063/1.1666171

\bibitem{Gergely:2002}
  L.~A.~Gergely,\\
  ``Linear Einstein equations and Kerr--Schild maps'',\\
  Class.\ Quant.\ Grav.\  {\bf 19} (2002) 2515\\
  doi:10.1088/0264-9381/19/9/313\\{}
  [gr-qc/0203101].



\bibitem{Santos}
Nilton O.~Santos,\\
``Newtonian Potential and the Complex Space'',\\
Lett. al Nuovo Cimento {\bf14}  (1975) 327--329.


   

\bibitem{Visser:2003}
M.~Visser,\\
  ``Physical wavelets: Lorentz covariant, singularity free, finite energy, zero action, localized solutions to the wave equation'',\\
  Phys.\ Lett.\ A {\bf 315} (2003) 219\\
  doi:10.1016/S0375-9601(03)01051-X\\{}
  [hep-th/0304081].

 

  \bibitem{DNJ:1996}
  M.~Demianski and E.T.~Newman,\\
  ``Combined Kerr--NUT solution of the Einstein field equations'',\\
  Bull. Acad. Pol. Sci., Ser. Sci. Math. Astron. Phys., {\bf14} (1996) 653--637.
  
   

\bibitem{Appell}
P Appell, \\
``Quelques remarques sur la th\'eorie des potentiels multiformes'',\\
(extrait d'une lettre adress\'ee \`a Mr. F. Klein)\\
P. Mathematische Annalen (Liepzig) {\bf 30} (1887) 155--156. \\
doi:10.1007/BF01564536


\end{thebibliography}
\end{document}